%% file: main.tex
\documentclass[USenglish,oneside,twocolumn]{article}

\usepackage[utf8]{inputenc}
\usepackage[big]{dgruyter_NEW}
 
\DOI{foobar}

\cclogo{\includegraphics{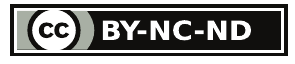}}

\usepackage{xspace}
\usepackage{xcolor}

\newcommand{\edits}[1]{{\textcolor{black}{#1}}}

\newcommand{\bheading}[1]{{\vspace{4pt}\noindent\textbf{#1}}}

\newcommand{\sysname}{{CoMPS}\xspace}

\begin{document}

  \author*[1]{Mona Wang}

  \author[2]{Anunay Kulshrestha}

  \author[3]{Liang Wang}

  \author[4]{Prateek Mittal}

  \affil[1]{Princeton University, E-mail: monaw@princeton.edu}

  \affil[2]{Princeton University, E-mail: anunayk@princeton.edu}

  \affil[3]{Princeton University, E-mail: lw19@princeton.edu}

  \affil[4]{Princeton University, E-mail: pmittal@princeton.edu}

\title{\huge Leveraging strategic connection migration-powered traffic splitting for privacy}

\runningtitle{Leveraging strategic connection migration-powered traffic splitting for privacy}


\input{abstract}

\keywords{website fingerprinting, traffic analysis, QUIC, WireGuard, multipath}

  \journalname{Proceedings on Privacy Enhancing Technologies}
\DOI{Editor to enter DOI}
  \startpage{1}
  \received{..}
  \revised{..}
  \accepted{..}

  \journalyear{..}
  \journalvolume{..}
  \journalissue{..}

\maketitle

\input{introduction}

\input{background}
\input{design}
\input{evaluation}

\input{results}

\input{future}

\section*{Acknowledgement}
We are grateful to Jonathan Mayer for insights that helped us refine this work. We are also grateful to anonymous reviewers and numerous HoTPETs 2021 attendees for their invaluable feedback. \edits{This work was supported by the National Science Foundation under grants CNS-1553437 and CNS-1704105, by DARPA under grant FA8750-19-C-0079, and in part, by the Ripple University Blockchain Research Initiative.}

\bibliographystyle{IEEEtranS}
\bibliography{bib}

\end{document}

%% file: abstract.tex
\begin{abstract}
{
Network-level adversaries have developed increasingly sophisticated techniques to surveil and control users' network traffic. In this paper, we exploit our observation that many encrypted protocol connections are no longer tied to device IP address (e.g., the connection migration feature in QUIC, or IP roaming in WireGuard and Mosh), due to the need for performance in a mobile-first world. We design and implement a novel framework, Connection Migration Powered Splitting (CoMPS), that utilizes these performance features for enhancing user privacy. With CoMPS, we can split traffic mid-session across network paths and heterogeneous network protocols. Such traffic splitting mitigates the ability of a network-level adversary to perform traffic analysis attacks by limiting the amount of traffic they can observe.
We use CoMPS to construct a website fingerprinting defense that is resilient against traffic analysis attacks by a powerful adaptive adversary in the open-world setting. We evaluate our system using both simulated splitting data and real-world traffic that is actively split using CoMPS. In our real-world experiments, CoMPS reduces the precision and recall of VarCNN to 29.9\% and 36.7\% respectively in the open-world setting with 100 monitored classes.
CoMPS is not only immediately deployable with any unaltered server that supports connection migration, but also incurs little overhead, decreasing throughput by only 5-20\%.
}
\end{abstract}

%% file: introduction.tex
\section{Introduction}
\label{sec:intro}
An individual's privacy today is closely tied to their network traffic. Malicious actors may devote vast resources to install network-control devices, either for commercial gain or to exert other forms of control.

The use of encrypted VPNs and Tor \cite{WireGuard_17, tor_04} for both privacy and censorship circumvention has skyrocketed in recent years as individuals counteract these trends. However, extensive research into traffic analysis techniques show that \emph{website fingerprinting} is feasible. Researchers have shown that encrypted Tor or VPN traffic can be fingerprinted, as each website exhibits unique packet-size and timing patterns that can be learned~\cite{frontglue, kfp, knnwfp, df}.

Many recent state-of-the-art defenses against website fingerprinting have shifted from obfuscating traffic to limiting the amount of information that the adversary can use to label a particular flow. Works like TrafficSliver \cite{trafficsliver}, which performs Tor-specific network-level splitting, and HyWF \cite{hywf}, which leverages multipath TCP, have demonstrated that splitting traffic across different network paths to limit the amount of information available to any network adversary is successful at thwarting many state-of-the-art website fingerprinting attacks. However, both these techniques face deployment challenges as they are specific to a particular proxy implementation or multipath TCP. 

In this paper, we exploit the observation that in order to improve performance on mobile clients, a growing number of encrypted protocols are no longer dependent on IP addresses and ports as identifiers, as in TCP. This feature is present in QUIC, WireGuard, and Mosh~\cite{quicietf, WireGuard_17, mosh}. This capability to switch network addresses is referred to as \emph{connection migration} in QUIC, and \emph{IP roaming} in WireGuard and Mosh.

We present a general framework that is deployable with any service supporting connection migration to foil powerful adversaries performing traffic analysis, surveillance, and selective censorship. Our system exploits connection migration features to support mid-session traffic splitting across heterogeneous paths and protocols.

Our Connection Migration Powered Splitting~(\sysname) framework is extremely flexible, as it enables clients to send packets over any network path or protocol available to the client \footnote{https://github.com/inspire-group/comps}. For instance, a multihomed device can utilize any of the network paths available to it. If the device has VPNs or proxies configured, the client can utilize those as separate \emph{\sysname paths} as well, so long as the original packet is delivered to the intended destination server and port. Previous traffic-splitting work has been limited to a particular protocol (such as Tor or multipath TCP), or limited to one network path with a co-operating router deployed~\cite{trafficsliver, mimiq, hywf}.

We demonstrate the resilience of website fingerprinting defenses constructed with \sysname. We evaluate \sysname as a website fingerprinting defense using both simulations over a dataset of QUIC website traces tunneled through WireGuard and a real-world dataset of around 30,000 QUIC website traces tunneled through WireGuard, split across \sysname. In our real-world splitting experiment, our results show that \sysname is highly effective in providing resilience against state-of-the-art website fingerprinting attacks.

We show that \sysname is also practical to deploy because of its low performance and implementation overhead.  We successfully deploy \sysname for unaltered QUIC, WireGuard, and Mosh servers, and demonstrate that the throughput impact for an effective splitting schedule is only around 5-6\% for WireGuard servers, around 10-20\% for QUIC servers, and around 4-5\% for Mosh servers.

Beyond website fingerprinting --- our primary focus in this work --- we also discuss how \sysname can enable a novel keyword-censorship circumvention method: the user can identify sensitive traffic (for instance, DNS requests, or a handshake containing the TLS SNI) to send over a high-latency path offered by a censorship circumvention mechanism, and migrate to the regular network path for the remainder of the session. We encourage the community to explore other privacy use cases of \sysname, and we will open source our framework to facilitate this. 


%% file: background.tex
\section{Background}
\label{sec:background}
We provide an overview of website fingerprinting, traffic splitting as a website fingerprinting defense, as well as \emph{connection migration} and \emph{IP roaming}.

\subsection{Website fingerprinting}
Website fingerprinting is a network traffic analysis attack aimed at determining which website a client is visiting from a stream of encrypted packets~\cite{df, knnwfp}. 

The attacker has the capability to eavesdrop packets sent on the network.  When a client encrypts and proxies their web traffic through VPN or Tor, the network adversary can only observe the resulting encrypted traffic to the tunneling service. Since they cannot decrypt the traffic nor observe the destination IP, the goal of the network adversary is to determine which website the client is visiting based solely on network metadata such as packet sizes and timing.

Many website fingerprinting attacks utilize features hand-crafted from packet size and timing data~\cite{og_fingerprinting_paper, wfp2011pachenko, cai2012_wfp, knnwfp, panchenko2016website,tiktok}. Random forest classifiers trained on these hand-crafted features can achieve a high accuracy on many website fingerprinting problems~\cite{kfp}. However, in recent years, the field has moved towards more powerful classifiers that can automate feature engineering using neural network architectures~\cite{df,varcnn, pfp,triplet}.

\subsection{Traffic splitting as a website fingerprinting defense}
Many website fingerprinting defenses other than traffic splitting involve the obfuscation of feature-rich metadata in encrypted web traces~\cite{wtfpad,csbuflo,walkietalkie,frontglue,tamaraw}. For instance, Tor cells are padded to a fixed size for this purpose~\cite{tor_04}. Despite the data overhead, this removes packet size information that a website fingerprinting adversary could otherwise use. Many other defenses also involve obfuscating packet timing, either by injecting ``dummy” packets or intentionally delaying packet delivery~\cite{wtfpad,frontglue}. The focus of this paper is on traffic splitting, which is an alternative or additional measure to obfuscation, and is built upon recent works that aim to reduce information that an attacker with a limited network perspective can learn~\cite{trafficsliver, hywf, mimiq}.
We examine recent website fingerprinting defenses built on traffic splitting and emphasize their limitations in comparison to \sysname. We summarize these differences in Table~\ref{tab:comparison-traffic-splitting}.

\begin{table*}[t]
\centering
\begin{tabular}{l|l|l|l}
                                        & Multiple paths & Requires client ISP support     & Protocol deployment    \\ \hline
TrafficSliver-Net & Yes                                  & No  & None (Modified Tor)      \\ \hline
HyWF             & Yes                                &  No  &   Low (Multipath TCP)       \\ \hline
MIMIQ             & No                               & Yes &  High (QUIC)                \\ \hline
\sysname             & Yes                              & No & High (Connection migration support) \\ 
\end{tabular}

\caption{Comparison of different traffic splitting systems for website fingerprinting resilience. MIMIQ's splitting occurs over a singular network path, meaning an adversary can see all client traffic, even if each subtrace is associated with a different IP address. MIMIQ also requires client ISP support. For other prior work, the protocol that performs the traffic merging is either custom (as in TrafficSliver) or not widely supported (as in Multipath TCP).}

\label{tab:comparison-traffic-splitting}
\end{table*}


\bheading{TrafficSliver.}
Previous work has explored splitting traffic across Tor circuits for improving Tor network performance~\cite{alsabah2013splitting,mtor,sangeetha_novel_2015}. Since then, others have proposed splitting traffic across Tor circuits for privacy, to defend against website fingerprinting by guard node operators~\cite{trafficsliver,yang_enhancing_2015,pennekamp_multipathing_2019}. In this model, the adversary either observes a single network path or operates a single guard node. We spotlight De la Cadena et al.'s proposed system, TrafficSliver, due to its extensive website fingerprinting evaluation. TrafficSliver, like other Tor circuit splitting proposals, relies on a change to the Tor protocol that allows the middle Tor node to merge traffic~\cite{trafficsliver,pennekamp_multipathing_2019}. Recent work has since built on these proposals, including creation of prototypes and further evaluation of their efficacy as website fingerprinting defenses~\cite{netsys-trafficsliver-demo, magnusson_evaluation_2021}. De la Cadena et al. also proposed splitting at the application layer, sending different HTTP requests down different Tor circuits, thus obviating the need to alter the Tor protocol to merge traffic at the middle node. However, the authors concluded that this latter method is not as resilient against website fingerprinting attacks~\cite{trafficsliver}.

Leveraging protocols with native support for connection migration, \sysname could perform this splitting without the need to alter Tor to merge traffic at the middle node. However, the connection migration protocols we examine rely on UDP, while the Tor protocol is tightly coupled with TCP connection state~\cite{tor_tcp_only}. If Tor builds out support for tunneling UDP, \sysname could split across Tor circuits as well.

\bheading{HyWF.}
Henri et al. propose a traffic fingerprinting defense that sends packets down different networks via Multipath TCP~\cite{hywf}. In this model, each adversary can only observe a single network path. Henri et al. also find that given the relatively low overhead of this method, it can be easily combined with other website fingerprinting defenses such as WTF-PAD and Walkie Talkie to further decrease website fingerprinting accuracy~\cite{wtfpad,walkietalkie}. \edits{We describe how CoMPS can easily replicate HyWF defenses without multipath TCP services in Section \ref{sec:comps-other}. This is especially useful as some popular middleboxes interfere with multipath TCP usage in practice~\cite{mptcp-wild}.} 

\bheading{MIMIQ.}
Govil et al. propose MIMIQ, where a router or ISP performs a form of traffic splitting on behalf of clients~\cite{mimiq}. The network device rebinds different IP addresses to a client throughout its connection. Unlike the previously discussed systems, MIMIQ traffic occurs over the same network path. Though MIMIQ investigates the capability of QUIC connection migration for privacy, its deployability is limited as the network device next to the client must cooperate. Since MIMIQ does not leverage multiple paths, it is also more vulnerable to flow correlation. Lastly, MIMIQ encodes a client identifier (although it can be somewhat obfuscated) in the IP address. The attacker could discern the mapping scheme used by MIMIQ to assign clients to IP addresses.

By contrast, \sysname utilizes other standard ways clients change or hide their IP address, such as using VPN or other anonymizing pathways, without needing to rely on a trusted network or closely deployed on-path ISP. In addition, \sysname supports splitting across different network paths, while all MIMIQ traffic traverses the same path. \sysname also supports other protocols that support \emph{IP roaming} such as WireGuard or Mosh.

\subsection{Connection migration and IP roaming}

For better mobility support, modern encrypted communication protocols may utilize a connection identifier that is distinct from IP addresses to identify the corresponding party. Examples include QUIC, WireGuard, and Mosh~\cite{quicietf, WireGuard_17, mosh}. In this section, we discuss mobility-related features in these three protocols, namely \emph{connection migration} in QUIC, and \emph{IP roaming} in WireGuard and Mosh.

\bheading{Connection migration in QUIC.}
QUIC is a UDP-based secure transport protocol that allows for multiplexing of application layer data streams. It was primarily designed to significantly improve HTTP performance in modern traffic patterns~\cite{quic}. The next-generation web protocol HTTP/3 will only support QUIC as a network-layer transport and not TCP~\cite{http3}. The QUIC transport also mandates encryption~\cite{quicietf}.

Many large services and CDN providers like Google, Facebook, Akamai, and Cloudflare support QUIC. The browsers Edge, Chrome, and Firefox all support upgrading connections to QUIC. In 2018, QUIC accounted for up to 9\% of Internet traffic, though this figure varied by vantage point~\cite{quic-wild}. In 2021, Cloudflare reported 13\% of the traffic it observes occurs over HTTP/3 (and thus, QUIC)~\cite{cloudflare_quic_v1}. We expect this number to continue increasing, especially once HTTP/3 becomes an RFC and as QUIC and HTTP/3 libraries stabilize~\cite{http3}.

QUIC uses a set of connection identifiers to uniquely identify an active connection across network changes. The foremost goal of this design choice is to provide smooth and fast migration when a device moves between different networks (such as from cellular to WiFi). Each QUIC connection usually has multiple valid connection IDs, which are communicated to the other endpoint through the encrypted tunnel once the connection is established. To reduce correlation between different traffic flows, QUIC necessitates the use of different connection IDs per each new network path~\cite{quicietf}.

When a QUIC client detects a network change, it first performs a round-trip \emph{path validation} to ensure the server is still reachable from the client. Neither party can exchange data packets until path validation has completed. After 2-way reachability has been established via path validation, servers may allow clients to freely migrate between paths that have been recently validated. Each time the connection is migrated onto the new network, congestion control parameters are reset. QUIC clients can also initiate connection migrations for other reasons, including network performance (if they sense degradation on one particular network path).

\bheading{IP roaming in WireGuard and Mosh.}
Other UDP-based encrypted protocols like WireGuard and Mosh support \emph{IP roaming}, which has similar properties to connection migration~\cite{WireGuard_17, mosh}.

WireGuard is a popular and flexible VPN protocol that has been incorporated into the Linux 5.6 kernel and backported into most major Linux distributions~\cite{salter_WireGuard_2020}. WireGuard is also the underlying protocol driving many commercial VPNs, including Mozilla VPN as well as Cloudflare's Warp VPN~\cite{mozilla_vpn, cloudflare_warp}. 
WireGuard performs some amount of congestion control per network path~\cite{WireGuard_17}.

Mosh is a remote terminal application similar to SSH that is intended for use by mobile clients~\cite{mosh}. Mosh is more resilient to network congestion, network drops, and network changes than SSH~\cite{mosh}.  Mosh is not built for high-volume transport, and generally adapts its frame-rate (and thus the amount of information it transmits) to network conditions~\cite{mosh}.

Unlike QUIC's \emph{connection migration}, endpoints that support \emph{IP roaming} are not expected to keep state about whether particular network paths have previously established 2-way reachability (via path validation) and do not explicitly reset congestion control parameters each time a packet is received on a new path. In \emph{IP roaming}, as long as an endpoint of the connection receives a valid, decryptable packet with a valid sequence number, that endpoint will accept the packet. 
WireGuard endpoints are manually configured with their peers' public keys, and in the case of Mosh, this key is established at the beginning of the session, bootstrapped over SSH.

%% file: design.tex
\section{\sysname: System design and implementation}
\label{sec:design}

With any service that supports IP roaming or connection migration, as long as each packet sent by the same client is received by the intended host, the packet can be delivered by any means, including different network paths on multihomed devices, encrypted VPNs, or other proxies. With this primitive, we can construct a framework to arbitrarily split ongoing sessions to any endpoint~(e.g., QUIC, Mosh, or WireGuard servers) that supports connection migration or IP roaming.

Currently, many network-control devices rely heavily on the source and destination IP address and port to identify TCP sessions. In general, \sysname enables users to split a continuous session across different IP addresses and ports, thus breaking this assumption. In the website fingerprinting context, we instantiate \sysname to foil network adversaries attempting to fingerprint traffic by limiting the amount of information they observe about a single session. This works against even powerful and adaptive adversaries.

In this section, we discuss the overall system design for \sysname and its core components that allow it to flexibly instantiate various types of network defenses. We detail how \sysname can be used to construct systems that can support a broad range of use cases against different adversaries.

 \begin{figure}[t]
    \centering
     \includegraphics[width=0.95\columnwidth]{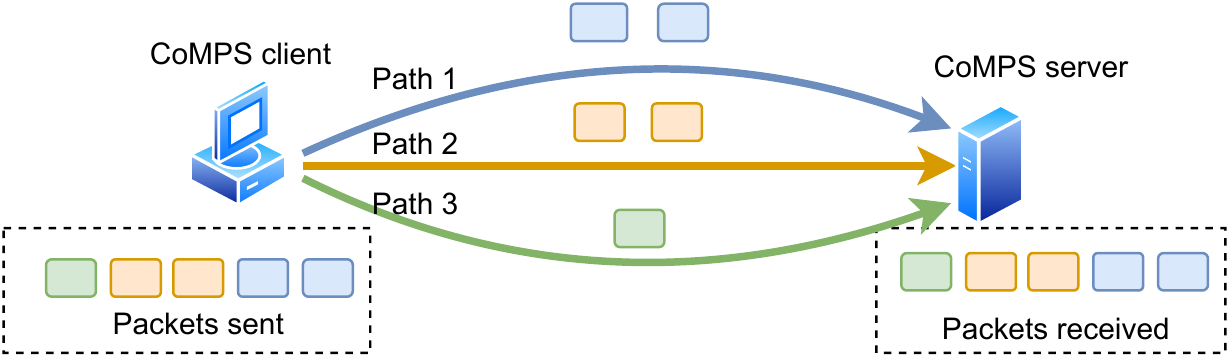}
    \caption{A path scheduler on the client sends packets down different CoMPS paths, which can encapsulate regular network paths as well as tunneling protocols or other forms of packet delivery.}
    \label{fig:design}
\end{figure}

\subsection{Design goals and adversary model}

The primary design goal for a \sysname deployment is to perform traffic splitting of a communication session. \sysname aims to achieve this functionality via client-side software: (1) without requiring any modification to servers, (2) without requiring any support from the client ISP, and (3) performing traffic splitting using widely-deployed protocols. Furthermore, \sysname is designed to enable traffic splitting across multiple network paths and tunneling protocols, which can support a range of use cases including website fingerprinting resilience and censorship circumvention.

Although precise attacker capabilities may vary depending on the use case, the \sysname framework generally focuses on a passive network-level adversary that performs traffic analysis of network traffic. \edits{Furthermore, the adversary is assumed to be limited in their perspective and can only observe traffic on one network path. A global adversary who sees all (or most) network paths is not considered. We also make no privacy claims in scenarios where an 
adversary observes multiple network paths and correlates traffic to the same client. \sysname focuses on the context of an honest client, and attacks that compromise client-side software are also out of scope. Finally, we assume that network traces are entirely encrypted (including DNS and SNI), consistent with prior work~\cite{wfquic}}.


\begin{figure*}[t]
\centering
    \includegraphics[width=0.95\textwidth]{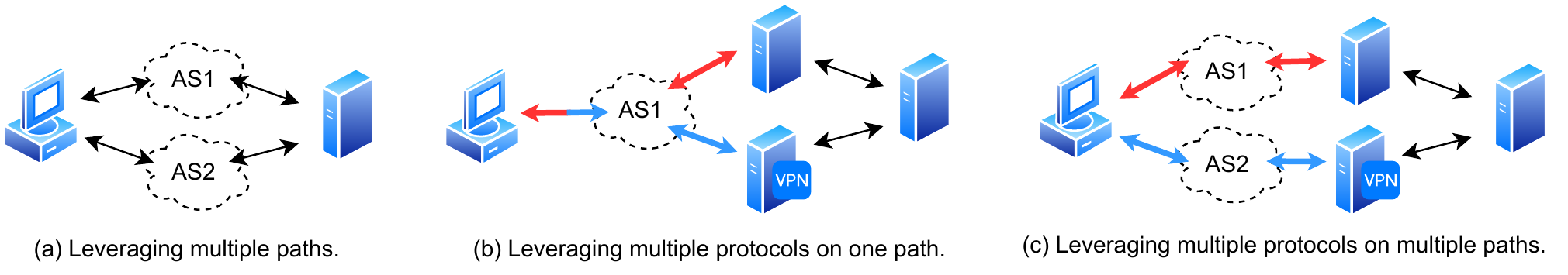}
    \caption{Demonstration of \sysname path components. In our threat model, we assume each adversary is limited to a single network position. We distinguish splitting between different network paths for multihomed devices as well as enabling splitting via different network protocols, via heterogeneous VPN or other encrypted tunnel protocols.}
    \label{fig:CoMPS-paths}
\end{figure*}

\subsection{Components of \sysname}
In any \sysname system, the client and server communicate over a protocol that supports connection migration or IP roaming. From there, the primary parameters that we can elect for our system are (1) the end server protocol, (2) different paths and protocols through which to deliver packets, and (3) the path scheduling algorithm that performs mid-session splitting and elects which path to send each packet across.

\bheading{(1) Connection migration-supporting server protocol.}
\sysname is deployable to any server supporting connection migration. In this paper, we examine QUIC, WireGuard, and Mosh. To differentiate between the three protocols, we refer to systems that connect to each as \sysname-over-QUIC, \sysname-over-WireGuard, and \sysname-over-Mosh, respectively.

\bheading{(2) Heterogeneous paths and tunneling protocols.}
As demonstrated in Figure \ref{fig:design}, a \emph{\sysname path} can be a regular network path, or a tunneled or proxying network protocol, like any VPN protocol, encrypted proxy, or SSH. The only requirement is that the packets from the client arrive at the same server through this path. The clients can choose, according to their own capability and limitations, how many and what type of \sysname paths to select. For instance, a multihomed client
can choose to send traffic over diverse network routing paths.
A non-multihomed client can elect to use different VPN protocols or encrypted tunneling protocols, so long as the packet arrives at the intended host.~\footnote{A protocol choice can include different instances of existing protocols with different configuration options that affect the traffic shape, e.g.,  WireGuard connections with varying maximum transmission unit (MTU) sizes.}

Choice of network paths can be limited by the device network (if the device is not multihomed), but choosing network tunnels is only limited by the preliminary effort to set up and maintain proxy or VPN servers. 

\begin{figure}[t]
    \centering

    \includegraphics[width=0.95\columnwidth]{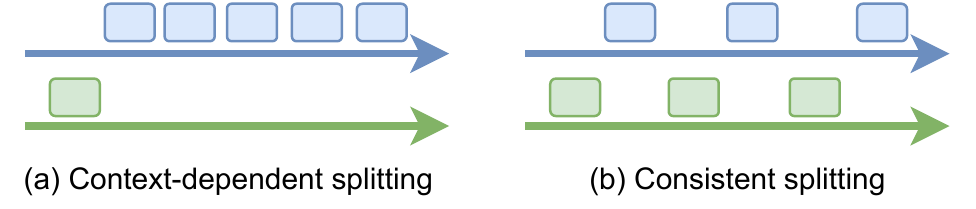}
    \caption{Demonstrating the difference between consistent and context-dependent splitting. In context-dependent splitting, the path switch may happen a constant number of times. In consistent splitting, the number of path changes is a function of the session length. In this example, we demonstrate round-robin switching, a straightforward consistent splitting strategy.}
    \label{fig:localized-consistent}
\end{figure}
 
\bheading{(3) Path scheduler.}
The path scheduler runs on the client device and provides a virtual network interface to the client program. The scheduler then routes packets down different network paths according to a predetermined path scheduling protocol. We consider two categories of path scheduling techniques: \textbf{consistent splitting} and \textbf{context-dependent splitting}. Consistent splitting is a regular and ongoing path-switching schedule; for instance, continuously switching paths after a set number of packets or milliseconds. Context-dependent splitting is more reactive, and is an occasional path-switch triggered by a particular network event.

The three types of \emph{consistent} splitting we highlight in this paper are \emph{round robin}, \emph{uniform random}, and \emph{weighted random}. Round robin deterministically alternates sending series of packets down each path. The \emph{uniform random} scheduler chooses each path uniformly at random, over which to send each series of packets. The \emph{weighted random} scheduler chooses a random weighted probability distribution for every connection. During traffic splitting, the scheduler samples each path from this probability distribution.

A \emph{context-dependent} splitting technique we examine is a path scheduler that prefers to send handshake packets to a proxy over an encrypted connection, and other packets over a regular connection. This example involves only one path switch, and the splitting is dependent on a particular context in the connection.

\subsection{Concrete use cases and traffic-splitting strategies}

In this paper, we primarily implement and evaluate \sysname instantiations and path-scheduling techniques for the concrete use case of defending against website fingerprinting. We also propose a \sysname technique to perform low-latency censorship circumvention. 

\bheading{\sysname as a fingerprinting defense.}
Recall that we describe website fingerprinting in Section \ref{sec:background}. For this use case, our threat model consists of an adaptive adversary who can observe encrypted traffic on ONE of the paths between the client and tunneling service, or between the client and server. From this perspective, they attempt to determine what websites the client is visiting.
They know which strategy the client is employing, and train their models with traffic that is defended using the same~strategy.

\sysname can replicate network topologies present in previous work on traffic splitting for website fingerprinting that utilize consistent path scheduling (constant and ongoing path switching), in addition to supporting a broader range of protocols. In this work, we will also implement instantiations using QUIC, WireGuard, and Mosh clients utilizing a consistent path-switching strategy, and evaluate their resilience against fingerprinting attacks and performance overhead.

\begin{figure}[t]
\centering
    \includegraphics[width=0.4\textwidth]{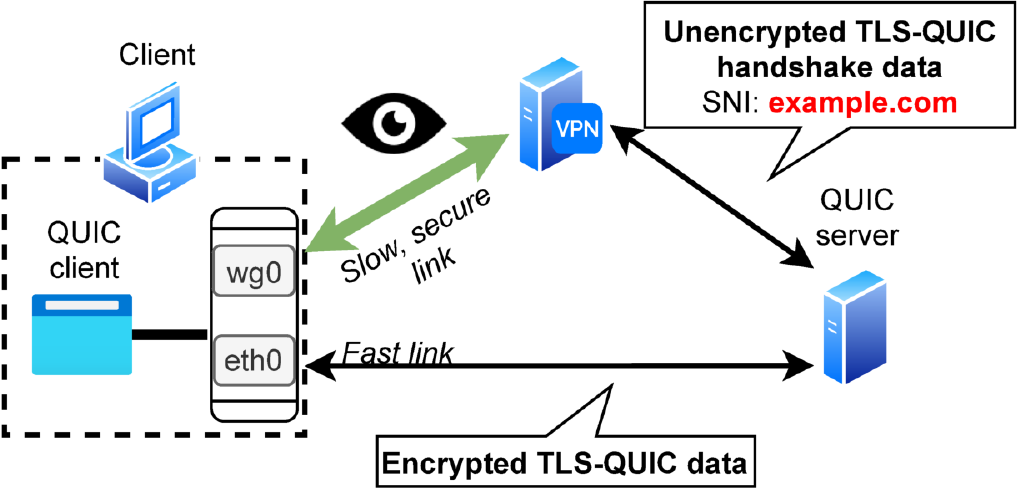}
     \caption{Using \sysname-over-QUIC to circumvent naive handshake-based censorship of TLS (with QUIC as the transport layer) connections with negligible overhead.} \label{fig:circumvent-design}
\end{figure}

\bheading{Circumventing handshake censorship.}
As encryption for both web and DNS connections has is slowly becoming the norm, ISPs increasingly use the unencrypted Server Name Indication (SNI) in the TLS handshake to identify connections to throttle or block~\cite{esni}. However, many reliable censorship circumvention transports (e.g., VPN, Tor, and Tor pluggable transports) are relatively high-latency, when compared to a native network connection.

Since TLS is integrated into QUIC, QUIC connections will also be susceptible to SNI-based censorship~\cite{quicietf}.
We design a simple \sysname topology to mitigate SNI-based connection identification with minimal overhead. We instantiate \sysname with two paths: a regular, non-VPN network path, and separate encrypted, VPN path (for instance, a WireGuard VPN). We employ the following context-dependent path scheduling strategy: the path scheduler sends any handshake packets over the VPN path. Once the handshake is complete, the session is migrated to the non-VPN path. The overhead of this technique is a single connection migration.

We illustrate this use case in Figure~\ref{fig:circumvent-design}. The design is similar to the concurrently developed BlindTLS, which uses TLS session resumption to switch network paths~\cite{blindtls}. Unlike BlindTLS, which focuses on TLS 1.2, our approach is independent of TLS protocol and works with TLS 1.3~\cite{blindtls}.

\subsection{Various \sysname topologies for website fingerprinting defense}

In this section, we describe the specific \sysname instantiations we evaluate in the paper: traffic splitting to any WireGuard VPN in Section~\ref{sec:comps-wg}, and traffic splitting to a QUIC server via WireGuard VPNs in Section~\ref{sec:comps-quic}.
In Section~\ref{sec:comps-other}, we also show that the \sysname framework subsumes existing traffic splitting designs, as well as provides for new types of traffic splitting systems with differing security properties.


\subsubsection{Multihoming to a WireGuard server}
\label{sec:comps-wg}
\sysname can enable arbitrary traffic splitting across network paths to any unaltered WireGuard VPN server. This enables the client to perform splitting with arbitrary network TCP and UDP network traffic that is tunneled through WireGuard. Here, we instantiate \sysname via the topology illustrated in Figure \ref{fig:CoMPS-paths}(a) to connect to a regular WireGuard VPN. In this scenario, our client is multihomed, and there is a passive adversary present on one of the upstream ISPs.

We evaluate the resilience of split WireGuard traffic on website fingerprinting attacks with both simulated and real data, as well as the overhead of performing splitting against a WireGuard server. The same approach can also be applied to Mosh.

\subsubsection{Traffic splitting to a QUIC server via VPNs}
\label{sec:comps-quic}
\sysname enables arbitrary traffic splitting across network paths to unaltered QUIC servers. However, to hide the handshake data (which, at the moment, reveals the destination server name), it is necessary to send this traffic through an encrypted tunnel. In this case, we instantiate \sysname with the topology in Figure \ref{fig:CoMPS-paths}(b) to connect to any QUIC server. In this scenario, the client does not need to be multihomed, in which case we consider an adversary that can only observe encrypted traffic on one of the paths between the client and VPN proxy. \edits{This defends against fingerprinting by ISPs further upstream from the user.}


We examine the effect of splitting QUIC-over-WireGuard traffic on website fingerprinting attacks. We also evaluate the overhead of splitting against a QUIC server that supports connection migration.

\subsubsection{Other \sysname topologies}
\label{sec:comps-other}

While we discuss the following extensions here for completeness, their evaluation is not our primary focus. Prior work has demonstrated the viability of these concepts~\cite{hywf, trafficsliver, mimiq} and the \sysname framework is able to realize their benefits in practice.

\bheading{Multihoming to a Tor bridge.}
If a device is multihomed, \sysname can also replicate the splitting techniques and defense topology presented in HyWF~\cite{hywf}. To achieve this, we instantiate CoMPS via the topology illustrated in Figure~\ref{fig:CoMPS-paths}(a) to connect to any Tor bridge that supports connection migration. 
Unlike HyWF, different \sysname instantiations can work without multipath TCP services, and also support heterogeneous protocols in addition to multiple network paths.

\bheading{Rotating source ports to foil naive network-control devices.}
MIMIQ proposes a system where an on-path network switch re-addresses traffic from a large pre-allocated pool of IP addresses available to the switch. This prevents a network adversary from naively correlating traffic across different IP addresses to the same connection~\cite{mimiq}. Since \sysname does not presuppose the existence of a friendly network switch that has access to a pool of IP addresses for address-masking, it cannot construct paths from new IP addresses.

We propose MIMIQ-lite, a \sysname instantiation where the client continually rotates its (IP, port) tuple by binding to new local source ports over which to send data. Although the split traffic may be correlatable via the IP address, in practice, network-control devices are heavily reliant on the (IP, port) tuple for connection identification. This would introduce some difficulty for existing (non-adaptive) adversaries/tools to correlate traffic occurring over constantly changing ports, even if the traffic does not utilize different IP addresses.

\bheading{Traffic splitting leveraging heterogeneous protocols.}
Unlike previous work in this area, \sysname also supports a potential strategy and topology where the client can switch between differing tunneling protocols (like WireGuard or OpenVPN) within a single session. This technique can also be combined with the techniques above --- that is, switching between source ports or network paths, to construct an even more resilient defense against network adversaries. Here, we can instantiate \sysname with the topology in Figure~\ref{fig:CoMPS-paths}(b) or \ref{fig:CoMPS-paths}(c), depending on whether the client is multihomed, to connect to any connection migration-supporting server.

The use of heterogeneous protocols can make it more challenging for the attacker to correlate flows from the same connection. Even an attacker who can monitor all the paths (which is not the primary attack model considered in this work) may not know for sure if they are produced by CoMPS's splitting or just two concurrent connections. If the adversary samples incorrect flows or omits relevant flows, this may cause additional false positives or false negatives. This technique is likely resilient against a naive (non-adaptive) attacker (i.e. one using a classifier, which is not trained on data split over heterogeneous protocols). Prior work on transfer learning between classifiers trained on different encrypted protocols has found that they generally perform quite poorly when applied to previously unseen encrypted protocols, even if the same data is being tunneled~\cite{wfquic}.

While these last few \sysname instantiations can raise the bar against non-adaptive adversaries, our primary evaluation focuses on splitting to WireGuard and QUIC servers, as described in Sections~\ref{sec:comps-wg} and \ref{sec:comps-quic}, which we show are resilient against website fingerprinting by an adaptive adversary.

\subsection{Implementation and deployment}
\label{sec:implementation}

\begin{figure}[t]
\centering
    \includegraphics[width=0.43\textwidth]{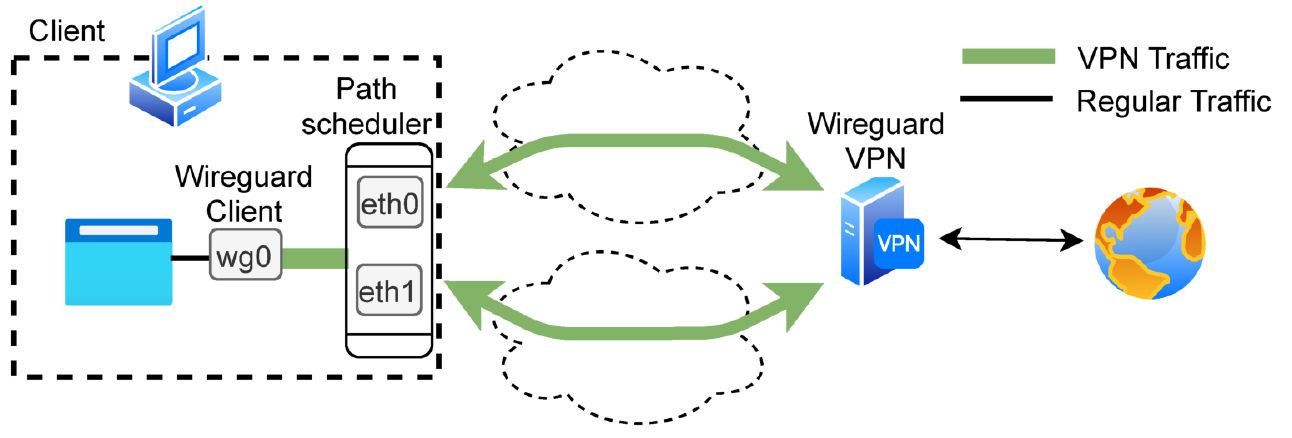}\\
    (a) CoMPS-over-WireGuard\\
    \includegraphics[width=0.43\textwidth]{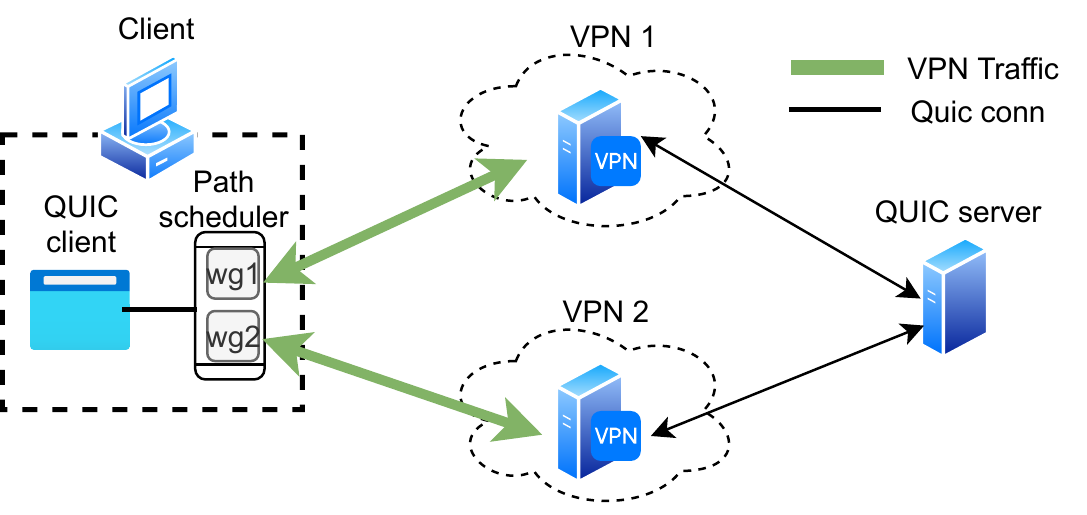}\\
    (b) CoMPS-over-QUIC\\
    \includegraphics[width=0.38\textwidth]{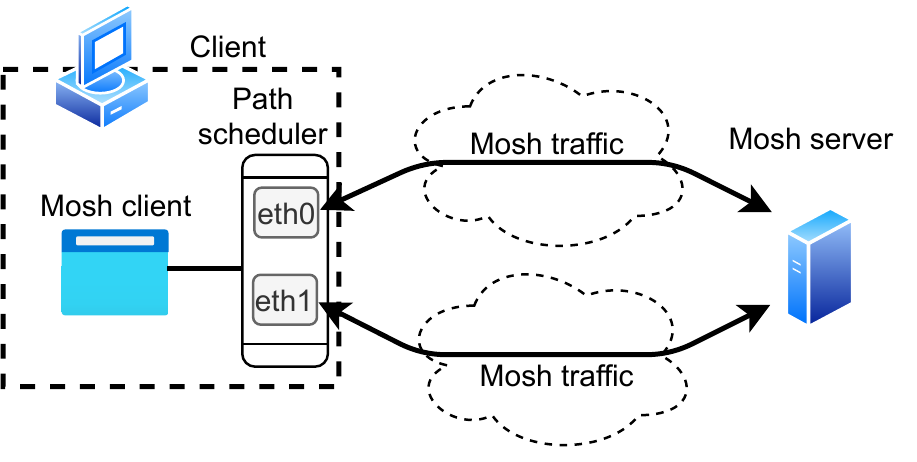} \\
    (c) CoMPS-over-Mosh
     \caption{Experimental setups for CoMPS with WireGuard, QUIC, and Mosh. Each path is abstracted to the client as a virtual network interface, to mimic how a client might intentionally swap between physical interfaces. } \label{fig:comps-setup}
\end{figure}

We now describe the \sysname deployment we use for performance experiments as well as data collection for subsequent evaluation as a website fingerprinting defense. We implement traffic splitting to a WireGuard server, as described in Section~\ref{sec:comps-wg} and demonstrated in Figure~\ref{fig:comps-setup}(a). We also implement splitting traffic to a QUIC server, as described in Section~\ref{sec:comps-quic} and demonstrated in Figure~\ref{fig:comps-setup}(b). Finally, we implement splitting to a Mosh server to show that \sysname can also work with other protocols that support connection migration.

\edits{Our client machine is located in New Jersey, although we run the real-world data collection from a virtual private server in Digital Ocean's NYC region.} Depending on which instantiation of \sysname a client deploys, they may need to first pre-allocate a number of encrypted tunnels, as in Figure~\ref{fig:CoMPS-paths}(b) or Figure~\ref{fig:CoMPS-paths}(c). \edits{In our experiment, we utilize WireGuard as our encrypted tunnel, deploying 2--3 VPN proxies for each network in addition to the server we are splitting traffic to. Our approach is compatible with any encrypted proxy, but we use WireGuard in our experiments. Depending on the experiment, these WireGuard proxies can either be local to the container, orchestrated via Docker Compose, or remote hosts, orchestrated via Ansible~\cite{docker, ansible}. }

\bheading{Emulating a Multihomed Setup.}
We emulated multihoming for client devices in our experiments by providing each device a number of virtual network interfaces. In practice, a user with a multihomed device would not need to do this.

\edits{We use Ansible and Docker Compose to orchestrate and manage multiple Docker containers on virtual private servers hosted on Digital Ocean~\cite{ansible, docker}. We use WireGuard to provide virtual network interfaces between containers on separate hosts. The arrows to the WireGuard VPN in Figure~\ref{fig:comps-setup}(a), and the arrows to the Mosh server in Figure~\ref{fig:comps-setup}(c) roughly correspond to WireGuard proxies in our test setup. The code is hosted at https://github.com/inspire-group/comps. }

\bheading{Implementing the path scheduler on the client.} 
Since each path is abstracted to the client as a virtual network interface, we run a process on the client side that continually changes the default network interface via \texttt{ip route}, based on a pre-determined path selection strategy. The path scheduler is thus not on-path for the packets, though it can passively observe them via packet capture. We focus here on consistent-splitting strategies for our website fingerprinting defense evaluation.

\bheading{\sysname-over-WireGuard.}
We instantiate \sysname with $n=3$ network paths in New York City, connecting to a WireGuard server in San Francisco. The WireGuard client can communicate with the WireGuard server over any of these paths. By attaching a container to the same network namespace as the WireGuard client, any and all traffic from that container is routed through this \sysname instantiation. The traffic is encrypted to the WireGuard server (travelling via one of the $n$ paths), and then forwarded to the intended destination. We use Selenium to orchestrate a Chrome browser to fetch various web sites over this network.

\bheading{\sysname-over-QUIC.}
For this experiment, we instantiate \sysname with several VPN proxies in New York City. This approach is compatible with any VPN protocol, but we use WireGuard VPNs here. In this case, we do not need a multihomed client. Since WireGuard VPN provides a virtual interface, it works well with the path scheduler. The QUIC client shares a network namespace with the WireGuard client which is able to send traffic through any of its peers. The QUIC client is then able to split traffic between remote WireGuard instances. It can also successfully fetch any remote QUIC resource so long as the underlying QUIC server implementation supports connection migration.

\edits{For our client, we used a forked version of Chromium’s simple QUIC client to perform connection migration~\cite{mimiq}, since their test client did not support connection migration.} At this time, Chromium's QUIC libraries do support connection migration, although whether it is used actively by the browser varies depending on client support~\cite{chromium_src, mozilla_neqo}. Generally, servers that utilize Google's QUIC libraries also support connection migration. Although connection migration is specified in the RFC, many other implementations of QUIC do not yet support it at this time. Large companies like Cloudflare and Facebook have already announced their intention to implement server-side connection migration~\cite{cloudflare_http3, facebook_quic}.

\bheading{\sysname-over-Mosh.}
For Mosh, we implemented a simple test deployment. We instantiated \sysname with 3 network paths, emulating a multihomed client. We ran an unaltered Mosh remote server located in New York City against which to test our setup, and on the client device, we installed and ran an unaltered Mosh client.

%% file: evaluation.tex
\section{Evaluation methodology}
\label{sec:evaluation}

We evaluate how the \sysname deployments detailed in Section~\ref{sec:design} perform under the conditions of an adaptive adversary performing website fingerprinting. In addition, we evaluate the relative overhead introduced by both QUIC connection migration and Mosh/WireGuard IP roaming.

\subsection{Threat model and defense evaluation}
We measure \sysname against a fully adaptive adversary (i.e. trained on data that is split using the same algorithm) using a neural network-based classifier in the \emph{open-world} setting, as described below. 

\bheading{The open-world setting.}
The study of website fingerprinting attacks (i.e. classifiers) and website fingerprinting defenses are designed to perform under either the \textit{open-world} model or the \textit{closed-world} model. In the closed-world model, the classification task is a multi-label classification problem where the classifier is trained on a set of \textit{n} web pages, and must label any future trace seen as one of these \textit{n} web pages. In the open-world model, the classifier is trained on a set of \textit{n} monitored web pages, as well as some large number of unlabelled unmonitored web pages. The classifier must first determine whether a trace belongs to the monitored set, and then label it. \edits{In this work, we use a train/test split of 9:1.}
We prefer evaluations under the open-world model as it describes an adversary who may encounter traffic belonging to new, unknown, or uninteresting web pages, which more realistically emulates the real-world traffic classification task.

\bheading{Metrics.} In the open-world model, the attacker must perform two classification tasks: first, a binary classification task to determine whether the trace belongs to a page in the monitored set, and if it is, they must perform an additional multi-class classification task to determine which of the monitored websites the trace belongs to. Metrics for evaluations in the open-world model tend to differ from the closed-world model (a standard multi-class classification task) due to this added complexity. For instance, an accuracy metric that only describes the ratio of correctly labelled traces to incorrectly labelled traces, even if weighted per-class, is not as useful since it does not distinguish between the different types of mistakes that such a classifier can make.

We use the open-world evaluation metrics provided by Wang et al.~\cite{wang2018optimizing}, which are more representative of classifier performance. A \emph{true positive} is when the classifier correctly labels a monitored trace. A \emph{wrong positive} is when the classifier labels a monitored trace as a different monitored trace. A \emph{false positive} is when the classifier labels a unmonitored trace as an monitored one. Recall is the rate of labelling a monitored trace as the correct monitored web page, and is equivalent to the true positive rate.

Instead of regular precision, we utilize \emph{r-precision}, presented by Wang et. al to more accurately capture the ratio of true positives to total positives~\cite{wang2018optimizing}. \emph{r-precision} scales the false positive rate by the natural base rate of unmonitored to monitored traces in the real world. It is defined as

$$\pi_r = \frac{\textrm{TPR}}{\textrm{TPR}+\textrm{WPR} + r \cdot \textrm{FPR}}$$

where $r$ is the expected base rate of unmonitored to monitored web page visits in the real world. As in Wang et. al's work~\cite{wang2018optimizing}, we use $r = 20$ in our evaluation. 

We also utilize the $F_1$ score, a more balanced depiction of classifier accuracy in the open-world setting. $F_1$ is the harmonic mean of the \emph{r-precision} and recall scores, calculated as

$$ F_1 = 2 \cdot \frac{\pi_r \cdot recall}{\pi_r + recall} $$

\bheading{Classifiers using neural networks.}
We evaluate \sysname against state-of-the-art classifiers built on neural network architectures: Deep Fingerprinting, p-FP, and VarCNN~\cite{df, pfp, varcnn}. Rather than using manually-extracted features, these classifiers borrow techniques from deep learning in computer vision by using convolutional neural networks to learn features from the raw packet traces.  We elect to use these classifiers to evaluate our splitting defense due to their high classification precision and recall (even against several state-of-the-art defenses)~\cite{df, pfp, varcnn}.

\bheading{Limitations of the open-world model.}
As explored in prior work, there are some practical limitations to the standard academic study of website fingerprinting~\cite{juarez_critical_2014}. For instance, website fingerprinting experiments generally assign labels to individual traces of automated browsers visiting the homepages of static websites. This hardly maps onto real client web-browsing behavior, where individuals load dynamic streams of content, many pages at a time, and will visit pages other than the homepages of websites.

We inherit these practical limitations in our evaluation. However, due to the expanding capabilities of commodity enterprise middleboxes and the growing willingness of state-backed network adversaries to perform more expensive forms of blocking, modelling powerful adversaries helps test the theoretical bounds of our defenses, so users can escape the proverbial ``cat-and-mouse'' game of network privacy and surveillance.

\subsection{Evaluation of simulated traffic splitting}

Collecting a real-world dataset of traffic splitting for each possible design and parameter choice in our splitting strategy is an expansive task. We thus first sought to determine the effect of path-switching strategy parameters on our classifier performance via simulated traffic splitting experiments. 

Using an existing dataset of QUIC-over-WireGuard traffic traces for open-world evaluation~\cite{wfquic}, we simulate a variety of splitting strategies across various parameter changes (number of paths, frequency of path switching, and path-switching strategy) while fixing the other parameters. This simulates the type of traffic an adversary might see in both the QUIC experimental setup (which splits QUIC traffic over different WireGuard proxies) as well as the WireGuard experimental setup (which splits any WireGuard traffic over different network paths).
By simulating the splitting, evaluating each set of parameters, and determining the ideal parameter choices for \sysname, we can then perform real-world traffic splitting and data collection in Section~\ref{sec:real-world-eval}. 

\subsubsection{Path-switching parameters}
\bheading{Number of paths.} This is the number of \emph{CoMPS paths} available to the client. Intuitively, splitting traffic across more paths means each potential adversary receives less information about the full network trace. However, realistically a user may not have many paths available to them, although this may change in the future. In our experiment, we vary the number of paths from 2 to 5.

\bheading{Frequency of path-switching.} This variable represents how often the scheduler must pick a new path. A higher switching frequency implies there are more variations to how a particular trace may be sliced. The most important tradeoff is that the performance may heavily degrade at high frequencies of path switching by confusing the protocol's congestion control algorithms. We evaluate performance at similar path-switching frequencies that are evaluated for resilience here. \edits{For the simulated data experiments, we primarily use packet-based evaluation (i.e. switching on the order of packets). }

\bheading{Path selection strategy.} At each path-switching junction, the scheduler must decide which path to send the next batch of packets. A simple strategy would be to deterministically \textit{round-robin} between each path. We also evaluate a \textit{uniform random} path selection strategy, where the path scheduler selects one of the paths uniformly at random. Lastly, we evaluate a \textit{weighted random} path selection strategy, where the probability distribution between paths is (1) non-uniform, and (2) newly chosen for each connection. For each connection, we sample the path probabilities from a Dirichlet distribution, a splitting technique previously proposed for splitting Tor traffic by De la Cadena et al.~\cite{trafficsliver}

\subsubsection{Simulating traffic splitting on a dataset}

We use the WireGuard dataset collated by Smith et al.~\cite{wfquic}, which contains both TCP and QUIC traces. \edits{Results on this dataset extend to both \sysname-over-WireGuard and \sysname-over-QUIC, as it provides traces of QUIC website traffic collected through WireGuard VPNs.} In total, there are 48,546 QUIC traces. The monitored set contains 100 domains, and the unmonitored set contains 16,182 traces. Monitored domains were sampled 100 times, and unmonitored domains were sampled 3 times \cite{wfquic}.

When we simulate splitting  for each trace, we break it into subtraces according to the number of \emph{\sysname paths}. Each subtrace is then considered as a separate trace in the training and test data.


\subsubsection{Combining \sysname with other website fingerprinting defenses}
\edits{\sysname, like other traffic-splitting techniques, can (and should) be combined with any website fingerprinting defense that utilizes obfuscation as its primary tactic.} We test combining \sysname with WTF-PAD, another low-latency website fingerprinting defense~\cite{wtfpad}. WTF-PAD is a natural fit since its primary mode of operation is to fill in packet timing gaps, caused by \sysname-like traffic splitting techniques.

\subsection{Evaluation and data collection of traffic splitting in the real world}
\label{sec:real-world-eval}

\edits{We evaluate \sysname on real-world data by collecting traces of network traffic split using our experimental setup described in Section~\ref{sec:design}.} We perform a website fingerprinting attack evaluation to show that \sysname provides effective resilience in real-world settings. Similar to the simulated experiments, our real-world data collection is evaluated under the open-world model with a fully adaptive adversary.

Using our WireGuard experimental setup depicted in Figure~\ref{fig:comps-setup}(b), we collect a dataset of website traces as a passive network adversary with a limited vantage point might observe them over this \sysname instantiation. The data collection setup uses 3 different network paths. The path scheduler performs the \emph{weighted random} strategy and switches every 100\,ms across 3 paths, which fits our desired performance constraints. We discuss the results from these initial experiments in Sections~\ref{sec:results-privacy} and \ref{sec:results-perf}. \edits{The dataset can be downloaded from https://github.com/m0namon/comps.}


We collect the traces of the top 10,000 websites according to Tranco, Majestic, and Alexa datasets. We combine and de-duplicate these domains, and select the domains that provide QUIC support. This process yielded 4,928 domains that could successfully load over HTTP/3.  We select 100 at random to be our monitored dataset, and use the remaining 4,828 websites as unmonitored traces. We fetch each monitored website 50 times, and each unmonitored website once. Since each fetch is split across 3 paths, each fetch produces three separate traces. This dataset is summarized in Table~\ref{tab:dataset}.

\begin{table}[t]
\centering
\begin{tabular}{l|l|l}
                                     & Monitored & Unmonitored \\ \hline
URLs          & 100        & 4,828         \\ \hline
Fetches per URL & 50        & 1           \\ \hline
\# Paths       & 3         & 2-3           \\ \hline
Total traces   &  15,000      & 14,484     \\ 
\end{tabular}


\begin{tabular}{l|l|l}
                                     & Monitored & Unmonitored \\ \hline
URLs           & 100        & 16,182         \\ \hline
Fetches per URL & 100        & 3           \\ \hline
Total traces   & 10,000       & 48,546        \\ 
\end{tabular}


\caption{Trace distribution for (\textbf{Upper}) our real-world dataset collected using \sysname~(split across 3 network paths), and (\textbf{Lower}) the QUIC-specific dataset from Smith et al.~\cite{wfquic}.}
\label{tab:dataset}
\end{table}

To collect a website trace, we use Selenium to instrument a Chrome browser to fetch the home page. In the meantime, several \texttt{tcpdump} processes capture the packets sent over each individual network interface. Each of these traces (over different network or WireGuard interfaces) are labelled as separate traces of that particular website. These traces are then converted into a vector of packet timing and sizes, which can then be consumed by website fingerprinting classifiers. \edits{As our threat model includes an adversary who observes one network path between the client and server, network traces from each link are treated as separate traces in the training stage. Recall that since \sysname does not employ Tor, packet sizes are also visible to adversaries.}

\subsection{Performance and overhead evaluation}
We describe our methods for evaluating the performance overhead for \sysname over QUIC, WireGuard, and Mosh.

We expect QUIC to incur higher overheads than WireGuard and Mosh due to its use of path validation and the resetting of congestion control parameters when switching to new paths~\cite{quicietf}. 
The use of single-path congestion control algorithms for multi-path network traffic may also affect all three deployments to varying degrees. We use the experimental setup described in Section~\ref{sec:design} to evaluate the practical overhead of all three protocols.

In all setups, we use 2 paths and the weighted random path scheduler. For QUIC and WireGuard \edits{we repeatedly fetch a remote resource until 1 GB of data is transferred. For Mosh, we repeatedly \texttt{cat} a 10\,MB file on a remote server. While each operation is running, the path scheduler is alternating the network path at a variable rate. Each experiment is repeated 10 times.}

%% file: results.tex
\begin{figure}
\centering
\includegraphics[height=0.25\textwidth]{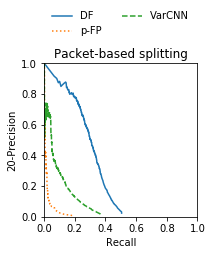} 
\includegraphics[height=0.25\textwidth]{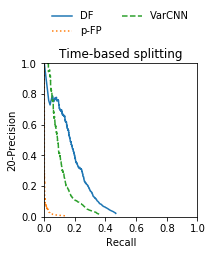} 
\caption{$r$-Precision v/s Recall curves from the simulated splitting experiments. On the left, we use packet-based splitting at a frequency of 50 packets. On the right, we use time-based splitting at a frequency of 100ms. We evaluate simulated \sysname against Deep Fingerprinting, VarCNN, and p-FP. For both experiments, we use the \emph{weighted random} strategy across 3 paths.}
\label{fig:simulated-defense}
\end{figure}

\begin{figure*}
\centering
\begin{tabular}{cccc}
 \includegraphics[height=0.23\textwidth]{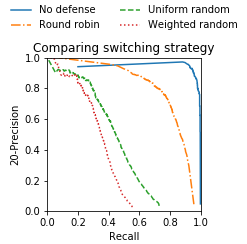} &   \includegraphics[height=0.23\textwidth]{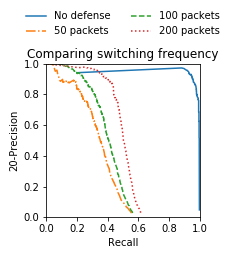} & 
  \includegraphics[height=0.23\textwidth]{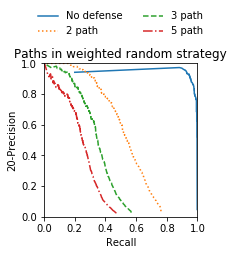} &
  \includegraphics[height=0.23\textwidth]{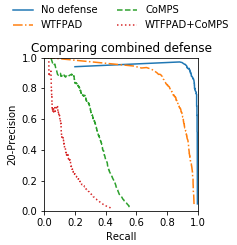} \\
  (a) Varying strategy &
(b) Varying switching frequency&

(c) Varying number of paths &
(d) Combining defenses\\[6pt]
\end{tabular}
\caption{$r$-Precision v/s Recall curves for the parameter experiments performed on simulated dataset. Our default setup used the \emph{weighted random} strategy, a path switching frequency of 50 packets, and three paths. (a) varies the switching strategy, (b) varies the switching frequency, and (c) varies the number of paths. (d) combines \sysname best performing strategy with WTF-PAD. \sysname is evaluated against the Deep Fingerprinting classifier \cite{df}. }
\label{fig:pr-curve}
\end{figure*}

\section{Resilience against website fingerprinting attacks}
\label{sec:results-privacy}

In this section, we discuss our evaluations of \sysname as a website fingerprinting defense, using simulated splitting as well as real-world splitting.

\begin{table}[t]
\centering

\begin{tabular}{ll|r|r|r}
\cline{3-5}
                                               & \multicolumn{1}{c|}{} & \multicolumn{1}{c|}{$r$-Precision} & \multicolumn{1}{c|}{Recall} & \multicolumn{1}{c}{$F_1$}   \\ \hline
\multicolumn{1}{l|}{\multirow{2}{*}{DF}}      & no defense            & 0.708                          & 0.993                       & 0.827 \\ \cline{2-5} 
\multicolumn{1}{l|}{}                         & defended              & 0.416                          & 0.329                       & 0.367                      \\ \hline \hline
\multicolumn{1}{l|}{\multirow{2}{*}{p-FP(C)}} & no defense            & 0.827                          & 0.801                       & 0.814                      \\ \cline{2-5} 
\multicolumn{1}{l|}{}                         & defended              & 0.415                          & 0.007                       & 0.014                      \\ \hline \hline
\multicolumn{1}{l|}{\multirow{2}{*}{VarCNN}}  & no defense            & 0.939                          & 0.953                       & 0.946                      \\ \cline{2-5} 
\multicolumn{1}{l|}{}                         & defended              & 0.462                          & 0.061                       & 0.107                      \\ \hline
\end{tabular}
\caption{Comparison of $r$-precision, recall, and $F_1$ on simulated splitting data across different classifiers. In this experiment, we use the \emph{weighted random} strategy across 3 paths and a switching frequency of 50 packets.}
\label{tab:diff-classifiers}
\end{table}

\subsection{Evaluation of WireGuard traffic splitting via simulated dataset}
\label{sec:results-simulated}
In the experiments discussed here, we used a path scheduler using the \emph{weighted random} strategy, switching every 50 packets, across 3 paths. In Section~\ref{sec:comps-param-results}, we vary each of these parameters individually.

\bheading{\sysname is effective against Deep Fingerprinting, p-FP, and VarCNN.}
As shown in Figure~\ref{fig:simulated-defense} and Table~\ref{tab:diff-classifiers}, \sysname performs well against multiple state-of-the-art website fingerprinting classifiers, significantly reducing both precision and recall compared to the baseline scenario.  
It performed especially well against the p-FP model, taking $r$-precision and recall down to 41.5\% and 0.7\% in the simulated splitting experiment from Table~\ref{tab:diff-classifiers}. The VarCNN classifier also achieved low recall, with precision and recall of 46.2\% and 6.1\%, and Deep Fingerprinting performed the best, at 41.6\% and 32.9\% precision and recall.

\bheading{Combining \sysname with WTF-PAD lowers website fingerprinting attack performance significantly.} In Figure~\ref{fig:pr-curve}(d), we demonstrate the effects of combining our best performing strategy with WTF-PAD. As also shown in Table~\ref{tab:combo_defense}, combining WTF-PAD with \sysname decreases classifier $r$-precision to 13.5\% and recall to 30.8\% on a simulated dataset. We thus show that not only is traffic splitting via \sysname an effective defense against a powerful and adaptive website fingerprinting adversary for WireGuard traffic, it can also be combined with other zero-latency website fingerprinting defenses to provide even stronger resilience.

\begin{table}[t]
\centering
\begin{tabular}{l|r|r|r}
\hline
 & \multicolumn{1}{c|}{\textbf{$r$-Precision}} & \multicolumn{1}{c|}{\textbf{Recall}} & \multicolumn{1}{c}{\textbf{$F_1$}} \\ \hline
No Defense & 0.708 & 0.993 & 0.827 \\ \hline
WTF-PAD & 0.348 & 0.957 & 0.510 \\ \hline
\sysname & 0.416 & 0.329 & 0.367 \\ \hline
WTF-PAD $+$ \sysname & 0.135 & 0.308 & 0.187 \\ \hline
\end{tabular}
\caption{Comparison of $r$-precision, recall, and $F_1$ across different defenses. The path scheduler used the \emph{weighted random} strategy across 3 paths with a switching frequency of 50 packets. These were evaluated with the Deep Fingerprinting classifier \cite{df}.}
\label{tab:combo_defense}
\end{table}

\begin{table}[t]
\centering
\begin{tabular}{c|c|c|r|r|r}
\hline
\textbf{Strategy} & \textbf{Batch Size} & \textbf{Paths} & \multicolumn{1}{c|}{\textbf{$r$-Precision}} & \multicolumn{1}{c|}{\textbf{Recall}} & \multicolumn{1}{c}{\textbf{$F_1$}} \\ \hline
\multicolumn{3}{c|}{\textbf{No Defense}} & 0.708 & 0.993 & 0.827   \\ \hline
RR & 50 & 3 & 0.743 & 0.774 & 0.758 \\ \hline
UR & 50 & 3 & 0.492 & 0.447 & 0.468 \\ \hline
WR & 50 & 3 & 0.416 & 0.329 & 0.367 \\ \hline 
\hline
WR & 50 & 2 & 0.582 & 0.511 & 0.544 \\ \hline
WR & 50 & 5 & 0.523 & 0.246 & 0.334 \\ \hline
\hline
WR & 200 & 3 & 0.681 & 0.477 & 0.561 \\ \hline
WR & 100 & 3 & 0.509 & 0.412 & 0.455 \\ \hline
\end{tabular}
\caption{Comparison of $r$-precision, recall, and $F_1$ across varying of path-switching strategies, path-switching frequencies, and path numbers. The Round Robin~(\textbf{RR}) strategy round-robins between the paths in a consistent order. The Uniform Random~(\textbf{UR}) strategy chooses a path uniformly at random. The Weighted Random~(\textbf{WR}) strategy samples the paths from a new non-uniform probability distribution for every connection. These were evaluated with the Deep Fingerprinting classifier \cite{df}.}
\label{fig:eval:merged}
\end{table}

\subsubsection{Evaluating \sysname parameters}
\label{sec:comps-param-results}
This section refers to results in Figure~\ref{fig:pr-curve} and Table~\ref{fig:eval:merged}, where we examine the impact of different parameters on \sysname's resilience against website fingerprinting.
Our default setup used the (1) \emph{weighted random} strategy, (2) a path switching frequency of 50 packets, and (3) three paths. For each experiment, we varied one of these parameters. Since Deep Fingerprinting performed the best in our experiment in Figure~\ref{fig:simulated-defense}, we use that classifier for our evaluations here.

\bheading{The weighted random strategy performs the best compared to the other path-scheduling strategies, as shown in Figure~\ref{fig:pr-curve}(a).} The deterministic round-robin strategy provides weaker resilience in comparison. Randomly choosing paths (as in the \emph{uniform random} strategy) improves the resilience, as the sampling pattern is no longer deterministic and thus more difficult for the attacker to learn. Finally, we find that the \emph{weighted random} strategy of choosing a new random distribution per-trace outperforms choosing paths uniformly at random.

\bheading{Higher path-switching frequencies improve resilience against website fingerprinting attacks, as shown in Figure~\ref{fig:pr-curve}(b).} A higher path-switching frequency increases the number of possible ways \sysname can split and sample the traffic, as there are more decision junctures throughout the packet trace. Many website traces are shorter than 1,000 packets; so a 200-packet switching frequency may only imply a few path switches during the connection. 



\bheading{Increasing number of paths improves resilience against website fingerprinting attacks.} Increasing the number of paths reduces the success of website fingerprinting classifiers, producing significant privacy gains even with just 2-3 paths. The experiment shown in Figure~\ref{fig:pr-curve}(c) was performed using the \emph{weighted random} strategy. 
While increasing the number of paths to 5 provides further gains in privacy, this design choice may not be practical for regular users, who are unlikely to have access to 5 separate network paths, even in the future. Since we want our system to be deployable, we primarily performed evaluations with 3 paths rather than 5. 

\edits{
\bheading{Splitting on time-based boundaries achieves similar results to packet-based boundaries.} 
To address the discrepancy between the time-based and packet-based boundaries used for the real-world and simulated experiments, respectively, we demonstrate that simulated splitting at an equivalent packet frequency (100\,ms and 50 packets) with the same weighted random strategy yields similar results, with the full PR curve illustrated in Figure~\ref{fig:simulated-defense}. With the DF classifier, splitting at 100\,ms with weighted random yields 28.1\% r-precision 25.1\% recall, respectively. This is very close to the performance of the classifier when trained on data that is split via the 50-packet boundary, which achieved 41.6\% r-precision and 32.9\% recall, respectively.
}


\subsection{Evaluation of WireGuard traffic splitting in the real world}


\bheading{Our real world experiment confirms \sysname’ ability to mitigate website fingerprinting.} 
We evaluate Deep Fingerprinting (DF), p-FP, and VarCNN on the real-world dataset split over \sysname using the \emph{weighted random} strategy, split across 3 paths, and a path-switching frequency of 100\,ms. This experiment was performed by switching at millisecond boundaries rather than a fixed number of packets due to the simplicity of the former implementation. Surprisingly, VarCNN performed better on the real-world dataset than on our dataset from simulated splitting, and Deep Fingerprinting performed worse on the real dataset. \edits{This might be partly due to VarCNN's design focus for low-data scenarios, as our real-world dataset is smaller than the simulated one.}

The differences show the importance of experimenting with multiple classifiers, and the collection of real-world data. Deep Fingerprinting achieved $r$-precision and recall of 12.4\% and 34.9\%, the p-FP model achieved $r$-precision and recall of 12.6\% and 12.1\%, and the VarCNN model performed the best, with an $r$-precision and recall of 29.9\% and 36.7\% respectively.

\begin{table}[t]
\centering

\begin{tabular}{l|r|r|r}
          & \textbf{$r$-Precision}       & \textbf{Recall}     & \textbf{$F_1$}    \\ \hline
\textbf{DF} & 0.124 & 0.349 & 0.183  \\ \hline
\textbf{p-FP(C)}    & 0.126 & 0.121 & 0.123  \\ \hline
\textbf{VarCNN}        & 0.299 & 0.367 & 0.329 \\ \hline
\end{tabular}



\caption{Comparison of $r$-precision, recall, and $F_1$ across real-world splitting data across different classifiers. Dataset collection is described in Section~\ref{sec:real-world-eval}.}
\label{tab:real-world}
\end{table}

\section{Overhead of traffic splitting}
\label{sec:results-perf}
We now summarize our results from performance experiments with \sysname. Recall that the performance experiments were performed by time-based boundaries rather than packet-based boundaries due to the simplicity of the implementation.

In our real-world data, the average throughput is approximately 500 packets/second. The 50 packet switching frequency from our simulated analysis in Section~\ref{sec:results-simulated} is thus approximately equivalent to 100\,ms path switching frequency for the real-world experiments. Our simulated results in Figure~\ref{fig:simulated-defense} generally show that time-based splitting at a frequency of 100\,ms achieves similar resilience against state-of-the-art neural networks to the equivalent packet-based splitting (every 50 packets).

\begin{figure*}[t]
\centering
\begin{tabular}{cccc}
    \includegraphics[width=0.25\textwidth]{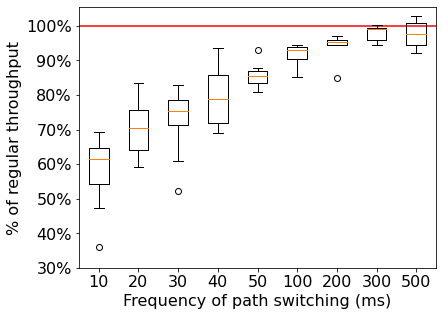}&
     \includegraphics[width=0.25\textwidth]{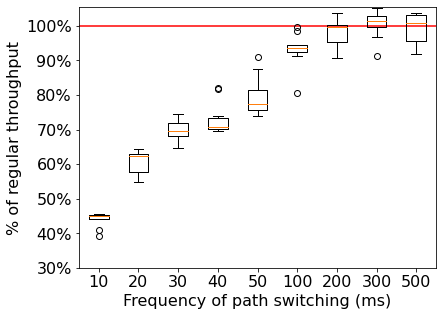} &
    \includegraphics[width=0.25\textwidth]{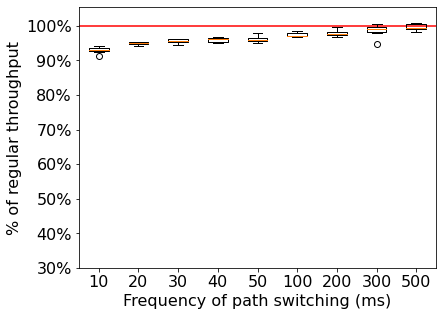} \\
    (a) \sysname-over-WireGuard &
    (b) \sysname-over-QUIC &
    (d) \sysname-over-Mosh \\
    
\end{tabular}
     \caption{Comparison of impact on throughput vs path-switching frequency for \sysname-over-WireGuard, \sysname-over-QUIC, and \sysname-over-Mosh (left-to-right). The red line is the average baseline throughput without path switching. The path scheduler randomly selects one of three paths at the specified frequency.}
     \label{fig:thruput}
\end{figure*}

\textbf{\sysname-over-WireGuard has low throughput overhead, with as little as 5–10\% overhead at reasonable switching frequencies, as shown in figure~\ref{fig:thruput}(a).} \sysname-over-WireGuard, the system we use to perform our real-world website fingerprinting evaluation, has quite low overhead at the 100\,ms switching frequency. We show that similar switching frequencies also provide reasonable resilience in our privacy evaluation in Section~\ref{sec:results-privacy}. Using faster switching frequencies (less than 40\,ms) leads to a reduction in throughput, and only offers marginal gains in our privacy evaluation.

\textbf{\sysname-over-QUIC has a 10–25\% throughput overhead at switching frequencies that provide resilience against website fingerprinting, as shown in Figure~\ref{fig:thruput}(b) and (c).} \edits{The larger amount of overhead from QUIC path migration is primarily due to path validations.} If path validations are cached, the performance impact would decrease for reasonable switching frequencies. At frequencies faster than switching every 30\,ms, the repeated need for path validation caused too many retransmission timeouts. To mitigate these performance effects, we encourage QUIC services to cache path validations.

\textbf{\sysname-over-Mosh has very little deterioration in performance overhead relative to switching frequency, as shown in Figure~\ref{fig:thruput}(d)}. Even at the unreasonably fast 10\,ms switching period, the average throughput overhead is less than 5\%. The low performance impact relative to QUIC or WireGuard is because Mosh generally does not transmit high-volume traffic, even if the remote terminal is updating quickly. Mosh adapts the frame rate of screen updates according to the expected connection RTT~\cite{mosh}.

\textbf{Mosh and WireGuard performance are less affected by traffic splitting than QUIC.}
Across Figure~\ref{fig:thruput}, we can compare how QUIC performance degrades more dramatically at higher frequencies, while WireGuard and Mosh performance is less impacted. We hypothesize that this is due to the congestion control window being reset on every path switch for QUIC.
The connection fails to exit the slow-start phase of the connection, leading to a drop-off in performance when the switching occurs too often. Nevertheless, at reasonable switching frequencies, the overhead is as little as 4-6\% for both WireGuard and Mosh and 10-20\% for QUIC. \edits{For comparison, experiments on TrafficSliver estimate a 20\% throughput overhead and WTF-PAD incurs a 60\% bandwidth overhead~\cite{trafficsliver, wtfpad}.}

%% file: future.tex
\section{Conclusion and future work}
\label{sec:future}

\bheading{\sysname is practical and effective to deploy as a website fingerprinting defense.} 
\sysname exploits various protocols' native capabilities for IP roaming and connection migration for network privacy. 
We successfully implement \sysname networks using QUIC, WireGuard, and Mosh; three protocols that support connection migration-like capabilities. We show that it is practical to constantly perform traffic splitting, and it introduces relatively low overhead. With a WireGuard-based \sysname defense, we can split any network traffic to the WireGuard VPN. 
In addition, QUIC~(and its connection migration capability) is continually gaining adoption. As Internet protocols develop to adapt to clients with dynamic IP addresses, \sysname-like splitting to servers will become even more feasible. In addition to its practicality, we also demonstrate that \sysname enhances resilience against fingerprinting attacks, even against powerful adaptive adversaries.

\subsection{Future work}
Due to the flexibility of \sysname, there is much room for further research in evaluating different ways to leverage connection migration for privacy.

\bheading{Testing \sysname for censorship circumvention.} 
We implement a simple prototype of our censorship circumvention use case (Section~\ref{sec:design}) using a QUIC-based CoMPS defense, which efficiently elides naive handshake censorship by sending handshake packets through an encrypted tunnel outside the censor's jurisdiction before migrating onto the regular network path. However, we did not thoroughly investigate the effectiveness of this technique in this work.

\bheading{Evaluating the tradeoff between traffic privacy and traffic visibility.} 
By splitting traffic across different network paths, \sysname (and other traffic-splitting systems) potentially expose portions of traffic to more network adversaries. Although each individual adversary may be able to learn less information about client traffic, the traffic is more likely to encounter more adversaries if it is split on multiple paths. This is in some ways similar to the trade-off that Tor users make when selecting the number of guard relays. This tradeoff has been examined in the Tor context~\cite{Dingledine14onefast,alsabah2013splitting}.

\bheading{Evaluating splitting across heterogeneous protocols.}
We did not fully explore the potential use of splitting a single traffic stream across heterogeneous protocols as a traffic analysis defense, which has never been explored in literature before, but is a possibility with \sysname. 
Prior work has found that website fingerprinting models trained on TCP-over-WireGuard traffic do not transfer well to QUIC-over-WireGuard traffic, and vice versa~\cite{wfquic}. When splitting traffic across different encrypted tunneling protocols, we can raise the bar for adversaries by forcing them to account for multiple protocols in their models since fingerprinting models do not transfer well across protocols.


\bheading{Using Tor circuits as \sysname paths.}
One avenue for future work is to examine the possibility of using Tor circuits as \emph{\sysname paths}. Unfortunately, Tor cannot tunnel UDP traffic natively as Tor's session model is currently tightly coupled with TCP~\cite{tor_tcp_only}. Thus, Tor cannot provide support for protocols like QUIC without an additional proxy, or a large engineering effort on the part of the Tor Project.

Many Tor developers and researchers have investigated adopting a UDP-based transport such as QUIC as a session-layer for Tor~\cite{datagramtor, udptor, turbotunnel, tor_quic_support_masters, quicTor}. Mathewson and Perry have also examined how switching to a UDP-based transport might affect their current security model~\cite{towards_analysis_udp_tor}. If Tor adopts a more flexible session layer, tunneling UDP packets could be possible. In the future, we would like to explore the implications of UDP-based (QUIC, for instance) traffic splitting over Tor for both censorship circumvention and resistance against traffic fingerprinting by a guard node. Given similar prior work examining traffic splitting over Tor for privacy, it is likely that \sysname would perform well~\cite{trafficsliver, yang_enhancing_2015,pennekamp_multipathing_2019}.

In summary, as surveillance of users' network communications continues to become more prevalent  and sophisticated, CoMPS offers a powerful new framework for resisting traffic analysis.
